\documentclass[11pt]{article}
\usepackage{moriond,epsfig}
\usepackage{graphics}
\usepackage{wrapfig}

\def\Journal#1#2#3#4{{#1} {\bf #2}, #3 (#4)}

\def\PRL{\em Phys. Rev. Lett.}
\def\PRD{{\em Phys. Rev.} D}

\def\ALL{\mathcal{A}_{LL}}
\def\pt{p_{\textrm{\scriptsize T}}}
\def\deltaG{\Delta G}
\def\deltag{\Delta g}

\def\be{\begin{equation}}
\def\ee{\end{equation}}
\def\bea{\begin{eqnarray}}
\def\eea{\end{eqnarray}}

\begin{document}
\vspace*{4cm}
\title{Recent results in high-energy longitudinal polarized
proton-proton collisions at $\sqrt{s} = $200GeV at RHIC}

\author{T. Sakuma for the RHIC Spin collaboration}

\address{Massachusetts Institute of Technology, Cambridge, MA 02139}

\maketitle\abstracts{We report on recent results of the longitudinal
double-spin asymmetry $A_{LL}$ from the STAR and PHENIX experiments.
Data were collected in longitudinally polarized proton-proton collisions
at a center-of-mass energy of 200 GeV. The results added new constraints
to the polarized gluon distribution function $\deltag(x)$ with the
probed $x$ range of $0.02 < x < 0.3$. The results lead to the
importance of probing small $x$.}

\section{Introduction}

The STAR and PHENIX experiments at the Relativistic Heavy-Ion Collider
(RHIC) at Brookhaven National Laboratory are carrying out a spin physics
program in polarized proton-proton collisions. One of the goals of the
RHIC-Spin program is to constrain the polarized gluon distribution
function $\deltag(x)$.

The longitudinal double-spin asymmetry $\ALL$ is defined by
\begin{equation}
\ALL = \frac{\sigma^{++} - \sigma^{+-}}{\sigma^{++} + \sigma^{+-}}
\end{equation}
where $\sigma^{++}(\sigma^{+-})$ is the cross section for the same
(opposite) colliding proton helicities. This asymmetry is directly
sensitive to $\deltag(x)$ through gluon-gluon and quark-gluon parton
level collisions. This asymmetry has been calculated using NLO
perturbative QCD for various parametrizations of $\deltag(x)$. We
measure $\ALL$ and test each parametrization by comparing the result
with theoretical calculations. In addition, our $\ALL$ results
contribute to global fits of the polarized parton distributions.

We test the validity of NLO perturbative QCD in the energy range of the
experiments using the cross sections for every final state for which we
measure $\ALL$.  We have found good agreement between the cross sections
and NLO perturbative QCD
calculations~\cite{starjetxsec}${}^{,}$~\cite{phenixpizeroxsec}${}^{,}$~\cite{phenixphotonxsec}.

\begin{wrapfigure}[17]{r}{3in}
\centering \scalebox{.4}{\includegraphics{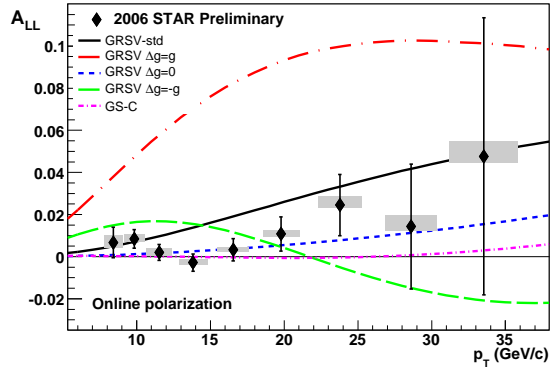}}

\caption{Longitudinal double-spin asymmetry $\ALL$ for inclusive jet
production at $\sqrt{s}=200$ GeV. The horizontal axis is jet $\pt$ for
particle jets. The statistical errors are shown by error
bars. The systematic uncertainties are shown by gray boxes. The curves
show NLO calculations based on the GRSV and GS-C polarized parton
distributions. \hspace*{12em}}
%$texttt{PARTICLE}
\label{fig:starjetsall}

\end{wrapfigure}

Since we observed the first polarized proton-proton collision at RHIC in
2002, we have measured $\ALL$ for various final states at a wide range
of rapidity at $\sqrt{s} = 62$ and $200$ Gev. In this paper, we show the
results of $\ALL$ measurements for inclusive jet production from the
STAR experiment and neutral pion production from the PHENIX experiment
at mid-rapidity at $\sqrt{s} = 200$ GeV from data collected during
the 2005 and 2006 RHIC runs. We show a quantitative comparison of $\ALL$
measurements to theoretical calculations.

\section{STAR inclusive jets}

The longitudinal double-spin asymmetry, $\ALL$, for inclusive jet
production was measured with the STAR detector. Jets used in this
analysis are reconstructed from tracks and neutral energy deposits and
defined by the midpoint-cone algorithm with a cone radius of 0.7. Tracks
were reconstructed with the time projection chamber (TPC) in a 0.5T
solenoidal magnetic field. The neutral energy deposits were measured
with the barrel electromagnetic calorimeter (BEMC), which covers a
pseudorapidity range of $|\eta| \le 1.0$ and full azimuth. The events
satisfied the jet patch (JP) trigger. This trigger requires a
coincidence between the east and west beam-beam counters and energy
deposits in a patch of calorimeter towers ($\Delta \eta \times \Delta
\phi = 1 \times 1$) greater than $8.3$ GeV. An integrated luminosity of
4.7$\textrm{pb}^{-1}$ was used in this analysis. The measured $\pt$ was
corrected to the particle level $\pt$ with a Pythia MC sample.

\begin{wrapfigure}[18]{r}{3in}
\centering 
\scalebox{.28}{\includegraphics{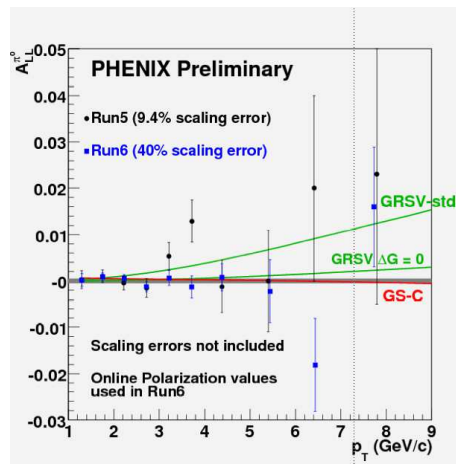}}

\caption{Longitudinal double-spin asymmetry $\ALL$ for neutral pion
production at $\sqrt{s}=200$ GeV versus $\pt$. The error bars show the
statistical errors. The curves show the NLO calculations based on the
GRSV and GS-C polarized parton distributions. \hspace*{11em}}

\label{fig:phenixpizeroall}

\end{wrapfigure}

Figure \ref{fig:starjetsall} shows the longitudinal double-spin
asymmetry, $\ALL$, for inclusive jet production together with NLO
calculations based on five different polarized parton distributions:
GRSV-max, GRSV-std, GRSV-zero,
GRSV-min~\cite{grsvstd}${}^{,}$~\cite{grsvmax}, and GS-C~\cite{gs}.
GRSV-std and GS-C were obtained from polarized DIS data but have very
different functional shapes. These polarized parton distributions have
very large uncertainties. GRSV-max, GRSV-zero, and GRSV-min are extreme
models.  The STAR inclusive jet $\ALL$ clearly excludes GRSV-max. A
quantitative discussion will be given below.

\section{PHENIX neutral pions}

Figure \ref{fig:phenixpizeroall} shows the longitudinal double-spin
asymmetry, $\ALL$, for neutral pions from the PHENIX experiment. Neutral
pions were reconstructed from two photons using the highly segmented
($\Delta \eta \times \Delta \phi = 0.01 \times 0.01$) electromagnetic
calorimeter. An integrated luminosity of 3.5$\textrm{pb}^{-1}$ was used
in the Run5 analysis and 7.5$\textrm{pb}^{-1}$ for Run6 analysis. The
final scale uncertainty for the Run6 analysis is expected to be around
10\% due to polarization uncertainties. The soft physics contribution
was estimated to be around 10\% at 2GeV/c and negligible for higher
$\pt$. The correction was made for the background $\ALL$.

\begin{wrapfigure}[18]{r}{2.6in}
\centering
\scalebox{.3}{\includegraphics{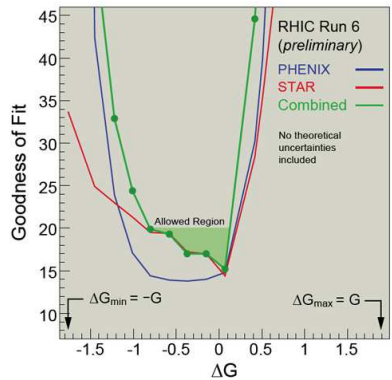}}

\caption{Constraints on the polarized gluon distribution, $\deltag(x)$
from Run 6 $\ALL$ measurement from STAR and PHENIX experiments. The
goodness of fit, $\chi^2$, as a function of $\deltaG$ was shown. The
range of $x$ probed by this analysis is $0.02 < x < 0.3$.\hspace*{19em}}

\label{fig:combine}

\end{wrapfigure}

\section{Quantitative theory comparison}

We quantified the impact of the $\ALL$ measurements on constraining the
gluon distribution $\deltag(x)$ within the GRSV framework. In Figure
\ref{fig:combine}, we compared the results with a series of $\deltag(x)$
assumptions which have the same functional shape as GRSV-std but have
different values of the first moment ranging from $\deltaG = -0.9$ to
$\deltaG = 0.7$ as well as four GRSV predictions: GRSV-max, GRSV-std,
GRSV-zero, GRSV-min.  The STAR and PHENIX results are consistent and
complementary and disfavor extreme gluon polarization scenarios of the
range of $x$ probed by this analysis~\cite{nsac} of $0.02 < x < 0.3$.

\section{Discussion}

The $\ALL$ comparison between data and NLO predictions was made within
the GRSV framework. GS-C has $\deltaG = 1.0$ and the $\ALL$ for GS-C is
consistent with both the STAR and PHENIX results. GS-C has highly
polarised gluons in the small $x$, less than the range of $x$ probed by
these measurements. To further understand the gluon's contribution to
proton spin, it is essential to measure $\ALL$ at small $x$.  The
measurements with the forward detectors are expected to play important
roles for understanding the small $x$ contribution.  In addition, the
500 GeV RHIC $\ALL$ measurements are expected to probe small x.

  The STAR and PHENIX experiments will continue to collect data from
polarized proton-proton collisions with wider energy range and wider
rapidity range in order to measure more precisely the polarized gluon
distributions, $\deltag(x)$, of the proton.

\section*{Acknowledgments}
I would like to thank the European Union "Marie Curie" Programme for the
financial support for my trip to attend the conference.

%$\section*{Appendix}

\end{document}